\documentclass[superscriptaddress,amsmath,amssymb,aps,prl,twocolumn,floatfix]{revtex4-2}

\usepackage{graphicx}
\usepackage{bm}
\usepackage[colorlinks, linkcolor=mycol, citecolor=mycol, urlcolor=mycol, breaklinks]{hyperref}
\usepackage{braket}
\usepackage{bbold}
\usepackage{xcolor} 
\usepackage{comment}
\usepackage{physics}
\usepackage{verbatim}
\usepackage{gensymb}
\usepackage{diagbox}

\newcommand{\mi}{ {\rm i} }

\definecolor{mycol}{RGB}{54,93,201}

\begin{document}

\title{Spin Squeezing with Itinerant Dipoles: A Case for Shallow Lattices}
\author{David Wellnitz}
\affiliation{JILA, National Institute of Standards and Technology and Department of Physics, University of Colorado, Boulder, Colorado, 80309, USA}
\affiliation{Center for Theory of Quantum Matter, University of Colorado, Boulder, Colorado, 80309, USA}
\author{Mikhail Mamaev}
\affiliation{JILA, National Institute of Standards and Technology and Department of Physics, University of Colorado, Boulder, Colorado, 80309, USA}
\affiliation{Center for Theory of Quantum Matter, University of Colorado, Boulder, Colorado, 80309, USA}
\author{Thomas Bilitewski}
\affiliation{Department of Physics, Oklahoma State University, Stillwater, Oklahoma 74078, USA}
\author{Ana Maria Rey}
\affiliation{JILA, National Institute of Standards and Technology and Department of Physics, University of Colorado, Boulder, Colorado, 80309, USA}
\affiliation{Center for Theory of Quantum Matter, University of Colorado, Boulder, Colorado, 80309, USA}

\date{\today}

\begin{abstract}
Entangled spin squeezed states  generated via  dipolar interactions in lattice models provide unique opportunities  for quantum enhanced sensing  and are now within reach of current experiments. A critical question in this context is which parameter regimes offer the best prospects under realistic conditions. Light scattering  in deep lattices can induce significant decoherence and strong Stark shifts, while shallow lattices face motional decoherence as a fundamental obstacle.
Here we analyze the interplay between motion and spin squeezing in itinerant fermionic dipoles in one dimensional chains using exact matrix product state simulations. We demonstrate that shallow lattices can achieve more than 5dB of squeezing, outperforming deep lattices by up to more than 3dB, even in the presence of low filling, loss and decoherence. We relate this finding to  SU(2)-symmetric superexchange interactions, which keep spins aligned and protect collective correlations. We show that the optimal regime is achieved for small repulsive off-site interactions, with a trade-off between maximal squeezing and optimal squeezing time.
\end{abstract}

\maketitle

Dipolar quantum gases made from polar molecules, Rydberg atoms, or magnetic atoms are emerging as promising platforms for near-term quantum technologies~\cite{baranov2012condensed,moses2017new,bohn2017cold,chomaz2022dipolar,browaeys2020many,morgado2021quantum}. These systems  are now routinely cooled to ultralow temperatures~\cite{ni2008high,lu2011strongly,aikawa2012bose,takekoshi2014ultracold,marco2019degenerate,phelps2020sub,son2020collisional,matsuda2020resonant,voges2020ultracold,guardado2021quench,schindewolf2022evaporation,stevenson2022ultracold},  and recently  pushed into a new  regime where individual particles can be controlled and measured using e.g.~quantum gas microscopes or optical tweezers~\cite{anderegg2019optical,zhang2020forming,burchesky2021rotational,christakis2022probing,holland2022ondemand,bao2022dipolar}. 

Taking advantage of these impressive developments defines a new frontier for quantum enhanced sensing. Of particular importance in this context is spin  squeezing~\cite{wineland1992squeezed,wineland1994squeezed}, which quantifies the reduction of uncertainty along a measurement axis due to quantum correlations~\cite{ma2011quantum} and also serves as a probe for many-body entanglement~\cite{ma2011quantum,sorensen2001many,pezze2018quantum}. 
It has been predicted that extensive spin  squeezing can be generated  in frozen dipoles trapped in deep optical lattices or optical tweezer arrays, where motional degrees of freedom are frozen and on-site collisions suppressed~\cite{yan2013observation,seesselberg2018extending,lepoutre2019out,anderegg2019optical,patscheider2020controlling,burchesky2021rotational,christakis2022probing,lin2022seconds,alaoui2022measuring,holland2022ondemand,bao2022dipolar}.
However,  the generation of spin squeezing via dipolar  interactions remains  an open challenge and so  far spin squeezing has been only  created  in atom and ion experiments with collective interactions~\cite{appel2009mesoscopic,sewell2012magnetic,hamley2012spin,muessel2014scalable,bohnet2014reduced,schmied2016bell,kruse2016improvement,cox2016deterministic,hosten2016measurement,bohnet2016quantum,baier2016extended,chalopin2018quantum,braverman2019near,bao2020spin,pedrozo2020entanglement}. This is  partially because  in frozen dipole setups~\cite{perlin2020spin} dephasing and dissipation  induced by off-resonant light scattering  and low filling fractions can significantly limit spin coherence times. In the context of polar molecules,  itinerant systems  confined in stacks of  2D pancakes have been considered as a promising alternative~\cite{bilitewski2021dynamical,li2022tunable}. However  at currently achievable temperatures, inelastic and lossy collisions in the pancakes~\cite{idziaszek2010universal,quemener2011universalities,croft2020unified,gregory2021moleculemolecule,cornish2022toward,liu2022bimolecular,bause2022ultracold} cannot be fully  suppressed, and have been observed to give rise to  motional dephasing and particle loss~\cite{li2022tunable}.

\begin{figure}
    \centering
    \includegraphics[width=\columnwidth]{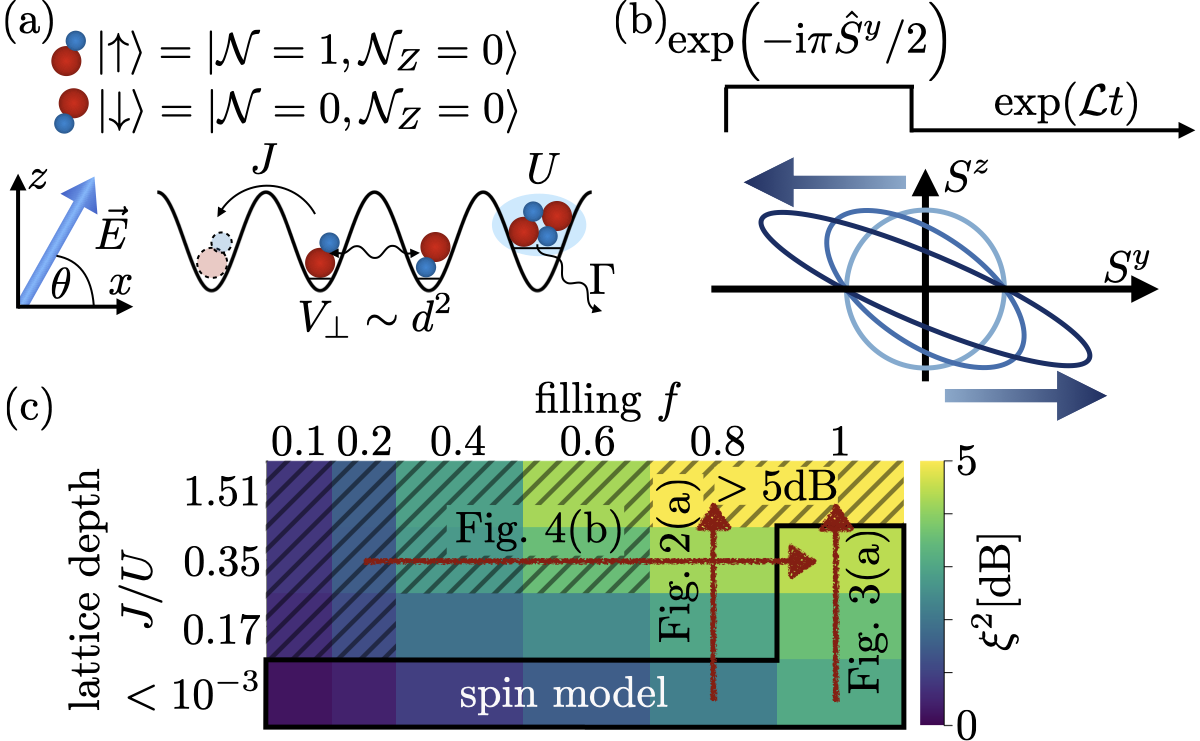}
    \caption{(a) Schematic of the  system: Fermionic dipoles   encoding a spin-1/2 degree of freedom in two internal levels (for molecules two  rotational states $\ket{\mathcal N, \mathcal N_Z}$ )  are loaded in a 1D chain.  The rotation axis is set by an external electromagnetic  field  at an angle $\theta$ to the lattice axis ($\vec E$ for electric dipoles). 
    Our model includes tunneling $J$, dipole interactions $V_\perp$, on-site interactions $U$, and two-body losses $\Gamma$.
    (b) A $\pi/2$ pulse prepares all dipoles in a superposition of both spin states, followed by free evolution with Liouvillian $\mathcal L$. Bottom: Schematic illustration of time evolution in the $S^y$-$S^z$-plane. With time, the interaction $V_\perp$ generates spin squeezing by shearing the quantum  noise distribution.
    (c) Maximal squeezing $\xi^2$ for $t < 10$ms versus filling fraction $f$ and lattice depth. In $x$ and $y/z$ directions $(V_{\mathrm{latt},x},V_\mathrm{latt,\perp})/E_\mathrm{R} = (3,3), (3,40), (5,40), (40,40)$ (top to bottom, see Supplemental for detailed parameters~\cite{SOM}). The black line indicates where the system can be approximated by a spin model. The striped area indicates where squeezing is growing past 10ms.
    }
    \label{fig:setup}
\end{figure}

Here, we study the exact quantum dynamics of fermionic itinerant  dipoles trapped in a 1D chain achievable  for example by imposing additional lattices along the 2D pancakes. Using matrix product states (MPS), and starting from a spin-coherent initial state, we  find that in all cases considered spin squeezing and coherence time are increased by reducing the lattice depth.  For  shallow lattices, particles remain itinerant~\cite{paz2016probing,fersterer2019dynamics} reducing  positional disorder at non-unit filling fractions, $0<f<1$, while undesirable lossy on-site collisions can be suppressed by the quantum Zeno effect~\cite{zhu2014suppressing,yan2013observation,sponselee2018dynamics}. Fig.~\ref{fig:setup}(c) summarizes these main results.
We further find that squeezing is enhanced when the signs of nearest neighbor dipole-dipole interactions and on-site interactions match, such that superexchange and dipole-dipole interactions add up. Smaller dipolar interactions give rise to larger squeezing, albeit at the cost of slower dynamics. We qualitatively explain these effects in a spin model valid for unit filling and sufficiently small tunneling.
We  also find that dephasing noise e.g.~due to differential lattice polarizability can be echoed away, as observed  in  recent experiments~\cite{li2022tunable}, without affecting squeezing dynamics. Even though we focus the analysis on polar molecules, our   predictions  apply to generic   itinerant 
fermionic  systems  featuring both contact and short-range off-site interactions.

\medskip

\textit{Model --- }We consider a 1D chain of fermionic dipoles trapped in an optical lattice with a spin-1/2 degree of freedom, which can for example be realized in the rotational states of molecules as $\ket{\uparrow} = \ket{\mathcal N=1,\mathcal N_Z=0}$ and $\ket{\downarrow} = \ket{\mathcal N=0,\mathcal N_Z=0}$, where a weak external field (electric field $\vec E$ or magnetic field $\vec B$) defines a preferred polarization axis [see Fig.~\ref{fig:setup}(a)]. The system is modeled as an extended Hubbard model with Hamiltonian~\cite{gorshkov2011quantum,gorshkov2011tunable} (see Supplemental~\cite{SOM} for a discussion of approximations)
\begin{align}
    \hat H &= \hat H_\mathrm{FH} + \hat H_\mathrm{dip}\, . \label{eq:hfull}
\end{align}

Here, $\hat H_\mathrm{FH}$ is the single-band Fermi-Hubbard Hamiltonian describing tunneling and on-site dipolar and contact interactions. It reads
\begin{align}
    \hat H_\mathrm{FH} &= -\sum_{j,\sigma} J_\sigma (\hat b_{j,\sigma}^\dagger \hat b_{j+1,\sigma} + h.c.) + U \sum_j \hat n_{j\uparrow} \hat n_{j\downarrow }\, , \label{eq:hhubbard}
\end{align}
with fermionic annihilation operators on site $j$ with spin $\sigma$, $\hat b_{j,\sigma}$, and number operators $\hat n_{j\sigma} = \hat b_{j,\sigma}^\dagger \hat b_{j,\sigma}$. The tunneling rates $J_\sigma$, and the contact interaction $U_\mathrm{contact}$ are controlled by the optical lattice depth, while the on-site dipolar interactions $U_\mathrm{dd}$ can be tuned via lattice depth, lattice anisotropy, and electric field ($U = U_\mathrm{contact} + U_\mathrm{dd}$). In general, a differential polarizability of the rotational states leads to spin-dependent tunneling rates, which can be tuned by the lattice polarization axis and are equal at a magic angle~\cite{neyenhuis2012anisotropic}. Close to zero field, the states $\ket \uparrow$ and $\ket \downarrow$ are spherically symmetric and have no induced dipole moments. Then, interactions between dipoles on different lattice sites are given by the dipolar exchange Hamiltonian $\hat H_\mathrm{dip} = \sum_{i>j} \frac{1}{\abs{j-i}^3} V_\perp \qty(\hat s_i^x \hat s_j^x + \hat s_i^y \hat s_j^y)$, and can be approximated as
\begin{align}
    \hat H_\mathrm{dip} = V_\perp \sum_j \qty(\hat s_j^x \hat s_{j+1}^x + \hat s_j^y \hat s_{j+1}^y) \label{eq:hdip} \, .
\end{align}
Here, the spin-operators $\hat s_j^\alpha = \hat \sigma_j^\alpha / 2$ with Pauli matrices $\hat \sigma^{x,y,z}$ are defined by $\hat \sigma_j^- = \hat b_{j,\downarrow}^\dagger \hat b_{j,\uparrow}$. In 1D, the $1/r^3$ tail of the interactions neglected in Eq.~\eqref{eq:hdip}  speeds up the squeezing dynamics, but the maximum attainable squeezing remains unchanged within numerical precision~\cite{SOM}. The interaction strength $V_\perp \propto [1 - 3\cos^2(\theta)]$ is controlled by the angle $\theta$ between the field and the orientation of the 1D  chain [Fig.~\ref{fig:setup}(a)].

We assume that particles are prepared in  their ground state $\ket \downarrow$ and uniformly distributed along the lattice such that each lattice site is occupied with probability $0 < f \leq 1$. Subsequently, a $\pi/2$-pulse prepares the dipoles in an $x$-polarized product state [see Fig.~\ref{fig:setup}(b)] with density matrix  $\hat \rho(t=0) = \bigotimes_j \hat \rho_j$ with
\begin{align}
    \hat \rho_j = (1-f)\ket{0}\bra{0}_j + f \ket{\rightarrow}\bra{\rightarrow}_j \, .
\end{align}
Here, $\ket 0$ is an empty lattice site and $\ket\rightarrow = (\ket\uparrow + \ket\downarrow)/\sqrt{2}$.

The system's dynamics is described by the Lindblad master equation
\begin{align}
    \partial_t \hat \rho &= \mathcal L \hat \rho = - \mi \qty[\hat H, \hat \rho] + \sum_j \mathcal D \qty[\hat L_j] \hat \rho \, , \label{eq:master} \\
    \mathcal D\qty[\hat L] \hat \rho &= 2 \hat L \hat \rho \hat L^\dagger - \hat L^\dagger \hat L \hat \rho - \hat \rho \hat L^\dagger \hat L \, .
\end{align}
On-site two-body losses e.g.~due to chemical reactions are described by Lindblad operators of the form $\hat L_j = \sqrt{\Gamma/2} \hat b_{j,\downarrow} \hat b_{j,\uparrow}$, where the loss rate $\Gamma$ increases with lattice depth~\cite{zhu2014suppressing,SOM}. We numerically simulate the dynamics of Eq.~\eqref{eq:master} by representing the vectorized density matrix as an infinite MPS directly in the thermodynamic limit, which we time-evolve with an infinite time evolving block-decimation algorithm~\cite{orus2008infinite,schollwock2011density,weimer2021simulation,SOM}. We measure squeezing by the Wineland squeezing parameter, which quantifies the precision gain in a Ramsey spectroscopy experiment~\cite{wineland1992squeezed,wineland1994squeezed,ma2011quantum}
\begin{align}
    \xi^2 = \frac{N(\Delta S_\perp)^2_\mathrm{min}}{\langle \vec S \rangle^2}\, . \label{eq:squeez_def}
\end{align}
Here, $ \vec S = (\hat S^\alpha)_{\alpha=x,y,z} = (\sum_j \hat s^\alpha_j)_{\alpha=x,y,z}$ is the Bloch vector, $4\langle \vec S \rangle^2/N^2=4\langle \hat S^x \rangle^2/N^2$ is the square of the contrast (since $\langle \hat S^y \rangle = \langle \hat S^z \rangle=0$) , and $(\Delta S_\perp)^2_\mathrm{min}$ is the minimal spin variance perpendicular to the Bloch vector [see illustration in Fig.~\ref{fig:setup}(b)]. We use a novel method to compute squeezing from infinite MPS directly in the thermodynamic limit, details of which are given in the Supplemental~\cite{SOM}. In all Figures, we use parameter values for fermionic KRb molecules, but we expect our findings to be relevant for arbitrary  fermionic dipoles.

\begin{figure}
    \centering
    \includegraphics[width=\columnwidth]{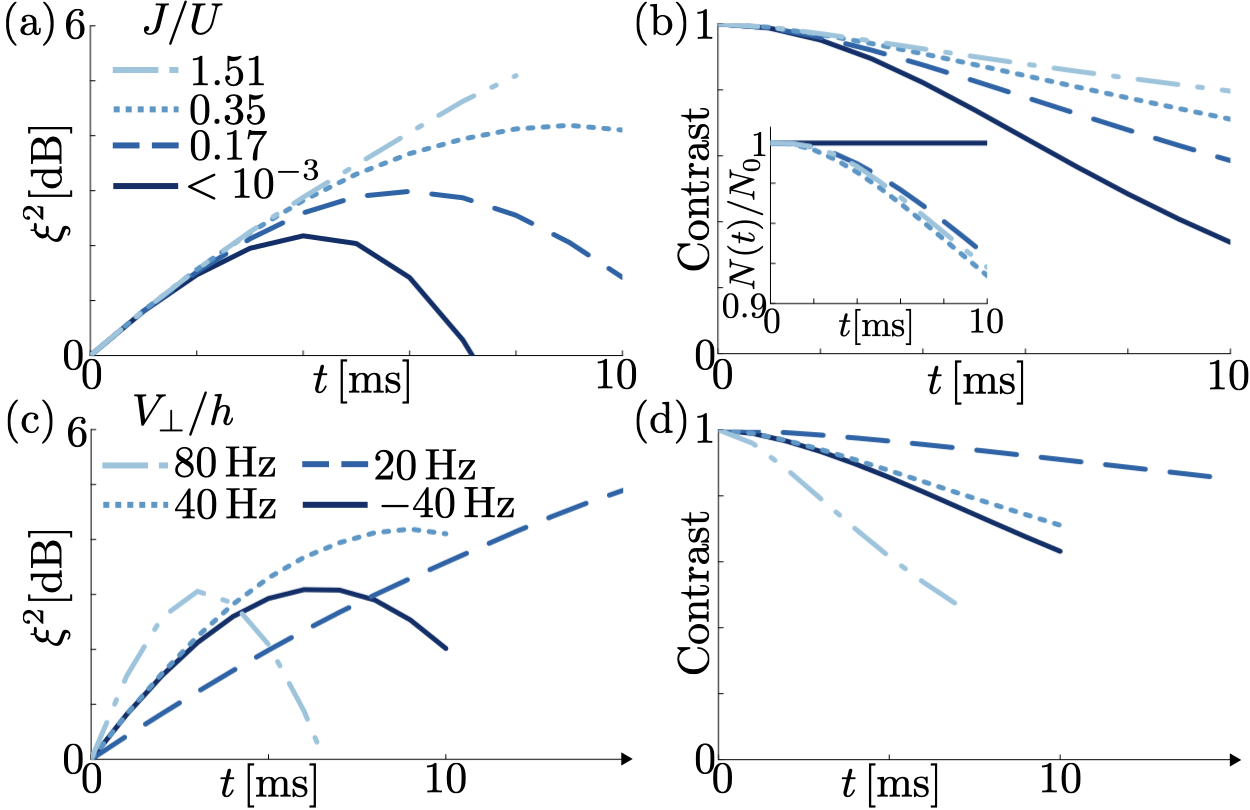}
    \caption{Full system dynamics. Time evolution of (a) squeezing $\xi^2$ and (b) contrast for various lattice depths [from light to dark, $(V_{\mathrm{latt},x},V_\mathrm{latt,\perp})/E_\mathrm{R} = (3,3), (3,40), (5,40), (40,40)$]. The inset shows the normalized molecule number $N(t)/N_0$. Parameters: $V_\perp/h = 40$Hz, $f=0.8$, $\Gamma=U_\mathrm{contact}/\hbar$.
    (c,d) Same as (a,b) for varying dipolar interaction strength $V_\perp$, while keeping the on-site interaction, $U$, and loss rate, $\Gamma$, fixed, and $(V_{\mathrm{latt},x},V_\mathrm{latt,\perp})/E_\mathrm{R} = (3,40)$. See Supplemental for lattice and MPS parameters~\cite{SOM}.}
    \label{fig:optimalregimes}
\end{figure}

\medskip

\begin{figure}
    \centering
    \includegraphics[width=\columnwidth]{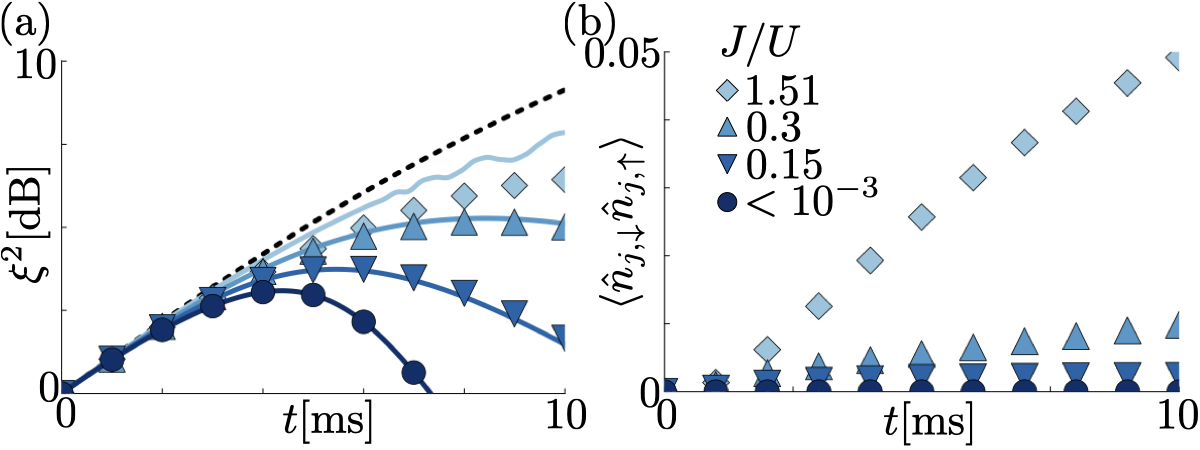}
    \caption{Full and spin model coherent dynamics at unit filling. Compared to Fig.~\ref{fig:optimalregimes} we set $f=1$ and $\Gamma=0$. Time evolution of (a) squeezing  $\xi^2$ and (b) doublon population $\langle \hat{n}_{j,\downarrow} \hat{n}_{j,\uparrow}\rangle$ for different lattice depths (as in Fig.~\ref{fig:optimalregimes}). Symbols represent results of the full dipolar Fermi-Hubbard model [Eq.~\eqref{eq:hfull}], continuous lines represent the spin model [Eq.~\eqref{eq:hsm}]. The black dashed line is the one axis twisting limit. $V_\perp/h = 40$Hz, $\Gamma = 0$.}
    \label{fig:coherent}
\end{figure}

\textit{Optimal parameter regimes --- }
We start by discussing the numerical findings for $J_\uparrow = J_\downarrow=J$, and give an analytical understanding in the following section. The maximal squeezing achieved within the first 10ms is shown in Fig.~\ref{fig:setup}(c) as a function of initial filling fraction and lattice depth. For all filling fractions, decreasing the lattice depth increases the squeezing. This is the main result of our paper and will be discussed in the remainder by considering time traces for parameters along the indicated arrows. We can see that while for deep lattices with frozen molecules ($J/U < 10^{-3}$) even for unit filling $f=1$ squeezing is limited to around 3dB, which constitutes a global maximum [see Fig.~\ref{fig:coherent}(a) below], shallow lattices can match and even out-perform these results for $f \gtrsim 0.4$. As will be discussed below in Fig.~\ref{fig:noise}(b), for such small filling fractions squeezing is limited by the evolution time and does not constitute a global maximum, as indicated by the grey striped area. This implies that if  times  longer than 10ms  were considered,    the results  would shift even more in favor of shallow lattices at low filling since the apparent saturation with lattice depth for small $f$ is limited only by the short-time growth, which we will show to be independent of lattice depth [Fig.~\ref{fig:optimalregimes}(a) and Fig.~\ref{fig:coherent}(a)].

Figs.~\ref{fig:optimalregimes}(a),(b) show the dynamics for different lattice depths at fixed filling $f=0.8$. Changing the lattice depth modifies both tunneling rate and on-site interaction, such that shallower lattice lead to a larger value of $J/U$. At short times, squeezing is generated at a rate independent of the lattice depth. For deep lattices, squeezing peaks at $\xi^2 \approx 2$dB. In contrast, for shallower lattices when molecules are itinerant, the growth persists longer, leading to larger maximal squeezing at later times. This is mirrored in the contrast decay in panel (b). While for deep lattices the contrast decays quickly, it remains much larger for shallower lattices. One might expect that the larger on-site loss rate and faster contrast decay and thus reduced Pauli blockade in deep lattices result in increased molecule loss. However, due to a combination of Zeno blockade and energetically suppressed doublon formation, the molecule loss is actually slowest in the deepest lattice [see Fig.~\ref{fig:optimalregimes}(b) inset]. As a consequence, losses remain below 10\% at all lattice depths.

Fig.~\ref{fig:optimalregimes}(c),(d) show the squeezing dynamics for a range of dipolar interaction strengths $V_\perp$. First, focusing on the results for $V_\perp/h = 40$Hz and $V_\perp/h = -40$Hz, it is clear that positive values of $V_\perp$ are preferable: The growth of squeezing persists longer and the coherence is maintained for longer. In order to observe the dependence on $\abs{V_\perp}$, consider the curves for $V_\perp/h = (20,40,80)$Hz. We find that increasing the interaction strength leads to a speed up of the dynamics, however at the cost of reducing the maximal squeezing. In order to achieve maximum squeezing, one thus wants to work with shallow lattices, and repulsive interactions. The optimal value of $\abs{V_\perp}$ is then determined by any dephasing mechanisms, which set a time scale limiting how slow the dynamics can be made.

\medskip

\textit{Analytical explanation: Spin model ---}
In order to provide insight into the underlying physics and  qualitatively understand the results discussed above, we now consider the limit $J\ll U$, $f=1$, and $\Gamma = 0$, where we can define an effective spin model [see Fig.~\ref{fig:setup}(c)]. Due to a combination of   Pauli and interaction  blockade mechanisms  at a small tunneling rate, molecules are essentially frozen in space. Each molecule can then be described as a localized spin~\cite{abrikosov1965electron}. The spins' interactions are governed by Eq.~\eqref{eq:hdip} and additional  super-exchange interactions from virtual hopping processes. The resulting Hamiltonian is an XXZ model given by:
\begin{align}
    \hat H_\mathrm{sm} &= V_\mathrm{sym,eff} \sum_j \vec s_j \vec s_{j+1} + V_{z,\mathrm{eff}} \sum_j \hat s_j^z \hat s_{j+1}^z  \label{eq:hsm}
\end{align}
with $V_\mathrm{sym,eff} = (4J^2/U) + V_\perp$ and $V_{z,\mathrm{eff}} = - V_\perp$ \cite{SOM}.

The XXZ model generates spin squeezing, which is largest for small negative values of $V_{z,\mathrm{eff}} / V_\mathrm{sym,eff}$~\cite{perlin2020spin}. The term proportional to $V_\mathrm{sym,eff}$ in Eq.~\eqref{eq:hsm} is SU(2) symmetric and thus cannot generate squeezing by itself, but favors spin alignment. It is largest for shallow lattices where one can reach larger values of $J/U$, and for $\mathrm{sgn}(V_\perp) = \mathrm{sgn}(U) = +1$. For these parameters the contrast is enhanced and the squeezing remains large (see Fig.~\ref{fig:optimalregimes}). Additionally decreasing $\abs{V_\perp}$ and thus $\abs{V_{z,\mathrm{eff}} / V_\mathrm{sym,eff}}$ further increases the maximal attainable squeezing. Finally, choosing $V_\perp > 0$ maximizes squeezing by ensuring $V_{z,\mathrm{eff}} / V_\mathrm{sym,eff} < 0$. Since $V_\mathrm{sym,eff}$ is SU(2) symmetric, the initial squeezing speed is determined solely by $\abs{V_{z,\mathrm{eff}}}$ = $\abs{V_\perp}$, independent of the lattice depth or the sign of $V_\perp$. It can be estimated by restricting dynamics to the fully symmetric manifold $\lvert\langle \vec S\rangle\rvert = N/2$, where the model reduces to the analytically solvable one axis twisting (OAT) model $\hat H = -\xi \hat S_z^2$ with $\xi = V_\perp / N$.

In Fig.~\ref{fig:coherent} we analyze the validity of the spin model for different lattice depths in absence of losses ($\Gamma = 0$). We find that, except for the shallowest lattice, the squeezing dynamics is well reproduced by the spin model [panel (a)]. For that case, while the initial growth rate is consistent with the OAT model, it overestimates squeezing at later times. A direct indicator of beyond spin model physics is the doublon population $\langle \hat n_{j\uparrow} \hat n_{j\downarrow} \rangle$ [panel (b)] which we find remains small  $\langle \hat n_{j\uparrow} \hat n_{j\downarrow} \rangle < 1\%$ at the time of maximal squeezing except for the shallowest lattice. For the latter, however, the doublon population becomes significantly larger $\langle \hat n_{j\uparrow} \hat n_{j\downarrow} \rangle \approx 5\%$ at the peak time, and the spin model breaks down. In the presence of losses, the small doublon population at moderate lattice depths also translates into losses, but they are small and typically less than 10\% of the initial molecules at the time of maximal squeezing~\cite{SOM}.

\medskip

\textit{Experimental considerations ---} 
Finally, we consider the impact of experimental imperfections on the generation of squeezing in Fig.~\ref{fig:noise}. Panel (a) shows the effect of spin-dependent tunneling rates, and panel (b) shows different filling fractions. Spin dephasing naturally arises due to the distinct polarizabilities of the spin states and the resulting state-dependent trapping potentials and tunneling rates $J_\uparrow \neq J_\downarrow$. Typical values for KRb at a lattice depth of $3E_R$ for the $\ket \uparrow$ state, are $J_\uparrow/h=153$Hz, $J_\downarrow/h=131$Hz \cite{SOM}. In Fig.~\ref{fig:noise}(a) we find that this leads to a reduction of spin squeezing from $\sim 4$dB to $\sim 2$dB. The tunnelling anisotropy can in principle be removed by a dynamical decoupling sequence, which effectively averages the tunneling rates of both states~\cite{li2022tunable,holland2022ondemand,bao2022dipolar}. Here, we consider a sequence of (infinitely fast) $X$ pulses $\mathrm{exp}(\mi \pi \hat S^x)$ spaced by a time $\tau$ and find that pulses with a pulse spacing of $\tau = 500$\textmu s are sufficient to almost fully recover the peak squeezing.

\begin{figure}
    \centering
    \includegraphics[width=\columnwidth]{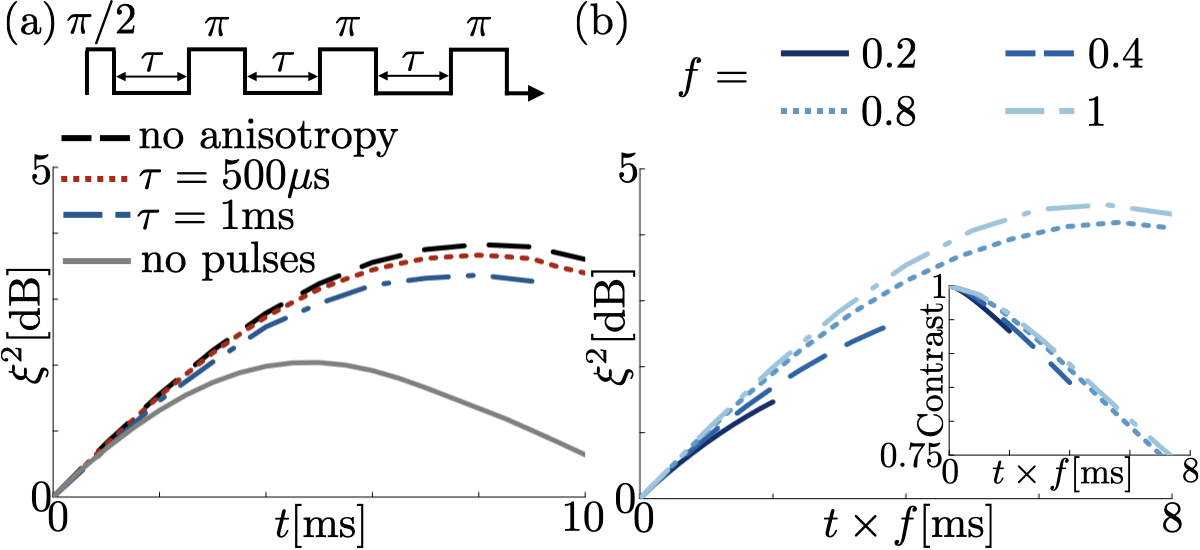}
    \caption{Spin squeezing $\xi^2$ in the presence of imperfections. (a) $X$-pulses with different pulse spacing $\tau$ protect against dephasing due to spin-dependent tunneling. We consider $J_\uparrow/h=153$Hz, $J_\downarrow/h= 131$Hz compared to the no anisotropy reference with $J_\downarrow/h = J_\uparrow/h = 142$Hz. Other parameters $f=0.8$, $U_\mathrm{contact}/h = 529$Hz, $U_\mathrm{dd}=0$, $\Gamma=U_\mathrm{contact}/\hbar$, $V_\perp/h= 40$Hz. (b) Dynamics for different filling fractions $f$. Inset shows the contrast decay. Parameters: $J/h=153$Hz, $U/h= 434$Hz, $\Gamma = 2\pi \times 512$s$^{-1}$, $V_\perp/h= 40$Hz.}
    \label{fig:noise}
\end{figure}

Panel (b) shows the squeezing dynamics at different filling fractions, corresponding to a horizontal cut through the diagram in Fig.~\ref{fig:setup}(c), versus time scaled by the initial filling fraction.  In experiments, the filling fraction is limited by the temperature of the gas before loading it into a lattice. We can compute the maximal achievable filling fraction by matching the entropy of free space gases to the entropy in the respective optical lattice.  While in 2015 experiments in optical lattices achieved filling fractions up to $f=0.25$~\cite{moses2015creation}, for $T/T_F = 0.3$ reported in Ref.~\cite{demarco2019degenerate}, theoretically filling fractions up to $f=0.9$ should be reachable
\footnote{A 3D harmonic oscillator at $T/T_F = 0.3$ has $S/(Nk_B) \approx 0.37$. A uniform filling fraction with $f=0.9$ has $S/(Nk_B) = - \ln(f) - (1-f) \ln(1-f)/f = 0.36$.}.

We observe a collapse of all curves when plotted as a function of the rescaled time $t \times f$. The slowdown of the dynamics is due to the reduction of average interactions $\propto f$. At later times, systems with lower filling fractions have reduced squeezing compared to the $f=1$ case, indicating that small filling fractions lead to a reduction of maximal attainable squeezing. Since the contrast is barely affected, the reduction in squeezing is due to an increase in the variance in Eq.~\eqref{eq:squeez_def}, which may be e.g.~due to enhanced motion at lower filling or disorder in the initial state.

Nevertheless, the most important reduction of the maximally reported spin squeezing for smaller filling fractions in Fig.~\ref{fig:setup}(c) is imposed  by the runtime of the dynamics, which here we set to 10ms, but will ultimately be limited   by additional sources of spin dephasing in an experiment.
Previous experiments in pancakes had coherence times limited by collisions~\cite{li2022tunable} which are already included in our analysis.  In a lattice, interaction-limited spin coherence times can be larger than $400$ms~\cite{christakis2022probing}, leading to negligible coherence loss on the 10ms time scales considered here~\cite{SOM}, thus supporting  the possibility to generate several dB squeezing in  current experiments.

\medskip

\textit{Conclusion ---}
In this paper, we have shown that spin squeezing is maximal in shallow lattices. While our study is focused on 1D due to the availability of exact numerical methods, we expect these results to extend to higher dimensions. In fact, the spin model arguments generalize directly to higher dimensions, and the better lattice connectivity in higher dimensions was shown to be beneficial in the case of deep lattices~\cite{perlin2020spin}.
Larger $\vec E$-fields and Floquet engineering provide additional tuning knobs, which can turn the XY into an XXZ model~\cite{gorshkov2011tunable,scholl2022microwave}, and thus further control $V_{z,\mathrm{eff}}$.
Additional density-spin interaction terms~\cite{gorshkov2011quantum} for large $\vec E$-fields may constitute an additional source of dephasing, which can however be removed by the pulse sequence discussed in Fig.~\ref{fig:noise}(a).

Furthermore, we emphasize that although in this paper we focused on KRb, we expect our results to generalize to other short-range interacting systems such as magnetic or Rydberg atoms. Indeed, the only necessary ingredients for our results are short-range interactions, the trapping in a tight-binding lattice, and the fermionic nature of the particles. The observed increase of squeezing with decreased interaction strength is particularly interesting for emerging experiments with magnetic atoms, which typically have much weaker interactions than molecules but longer coherence times~\cite{chomaz2022dipolar}. It is an interesting prospect to consider if our results can be further extended to bosons, and how they ultimately  translate  when pushing to even shallower lattices when corrections to the Fermi-Hubbard model become important~\cite{wall2017microscopic,hughes2022accuracy}.

\bigskip

 We acknowledge careful review of this manuscript and  useful  comments  from Jun-Ru Li and Jacob Higgins. We thank Johannes Schachenmayer for valuable contributions to the MPS code, which makes use of the intelligent tensor library (ITensor)~\cite{itensor}.
The work  is supported by the AFOSR MURI,  the ARO single investigator award W911NF-19-1-0210,   the  NSF JILA-PFC PHY-1734006 grants, NSF QLCI-2016244 grants, by the DOE Quantum Systems Accelerator (QSA) grant and by NIST.

\bibliography{main}

\clearpage
\newpage

\appendix

\setcounter{equation}{0}
\setcounter{figure}{0}
\setcounter{table}{0}
\setcounter{secnumdepth}{2}
\makeatletter
\renewcommand{\thefigure}{S\arabic{figure}}

\begin{widetext}

\section{Hamiltonian Derivation and Parameters} \label{app:parameters}

\begin{table}[b]
    \centering
    \begin{tabular}{c|cccc}
        Lattice depth [$E_R$] & $J/h$ & $J_\perp/h$ &  $U_\mathrm{contact}/h$ & $U_\mathrm{dd}/h$ \\
        \hline
        $3 \times 3 \times 3$ & 153Hz & 153Hz & 101Hz & 0\\
        $3 \times 40 \times 40$ & 153Hz & 0.15Hz & 512Hz & -78Hz \\
        $(3/3.6) \times 40 \times 40$ & 131Hz & 0.15Hz & 529Hz \\
        $5 \times 40 \times 40$ & 90.9Hz & 0.15Hz & 614Hz & -73Hz\\
        $40 \times 40 \times 40$ & 0.15Hz & 0.15Hz & 1151Hz & 0
    \end{tabular}
    \caption{Lattice parameters for KRb with lattice laser wavelength $1064$nm (532nm lattice spacing). The case $3/3.6$ refers to the anisotropic lattice with depth $3E_R$ for molecules in $\ket \uparrow$ and $3.6E_R$ for molecules in $\ket\downarrow$. $U_\mathrm{dd}$ is here given for an angle of $47.4\degree$, which matches $V_\perp/h = 40$Hz. $J_\perp$ is the tunneling rate perpendicular to the $x$ direction, which we set to zero for a 1D geometry.}
    \label{tab:lattice_params}
\end{table}

In the tight binding limit, we write the Hamiltonian in a basis set of localized Wannier orbitals on site $j$, $\psi_j(x) = \psi(x-ja)$, with $a$ the lattice spacing. The derivation of the Fermi-Hubbard model for an optical lattice is well known, and can e.g.~be found in Refs.~\cite{bloch2008many,wall2017microscopic}. Here, we only repeat some main results for convenience and point out some important features for molecules. The tunneling rate is generally given by
\begin{align}
    J = - \frac{\hbar^2}{2 m} \int dx \, \psi^*(x) \partial_x^2 \psi(x-a) + V_\mathrm{latt} \int dx  \, \psi^*(x) \psi(x-a) \sin^2(\pi x/a) \, ,
\end{align}
with $m$ the particle mass. The on-site interactions are the sum of contact and dipole-dipole interactions $U=U_\mathrm{contact} + U_\mathrm{dd}$. The contact interactions can be computed from the overlap integral and the $s$-wave scattering length as
\begin{align}
    U_\mathrm{contact} &= \frac{4\pi \hbar^2 \Re(a_\mathrm{sc})}{m} \int d^3 r\,  \abs{\psi(\vec r)}^4 \, .
\end{align}
The values for the lattice depths considered in the paper are given in Tab.~\ref{tab:lattice_params}.

\subsubsection{Differential spin polarizabilities}
If the lattice depth experienced by  the two spin states is different, the spin states will have different Wannier orbitals $\psi_\uparrow(x)$ and $\psi_\downarrow(x)$, and thus distinct tunneling rates $J_\uparrow \neq J_\downarrow$. The overlap integral needs to be adapted to $U_\mathrm{contact} = \Re(a_\mathrm{sc}) \int d^3 r \, \abs{\psi_\uparrow(\vec r)}^2 \abs{\psi_\downarrow(\vec r)}^2$. In particular, for KRb, at a $90\degree$ angle between the lattice polarization and the $\vec E$-field, the lattice for the $\ket\downarrow$ state is approximately 20\% deeper than for the $\ket\uparrow$ state~\cite{neyenhuis2012anisotropic}, leading to the values shown in Tab.~\ref{tab:lattice_params}.

\subsubsection{Dipole-dipole interactions}

In this section we give a concise summary of the dipole-dipole interactions at weak field closely following Ref.~\cite{gorshkov2011quantum}. The rotational level structure of a single molecule in its electronic and vibrational ground state is described by the Hamiltonian
\begin{align}
    \hat H = B \hat N^2 - E \hat d_0 \, ,
\end{align}
where $B$ is the rotational constant, $\hat N$ is the angular momentum operator, $E$ is the electric field, and $\hat d_0 = d\, Z$ is the dipole operator in the direction of the electric field $Z$ with the permanent dipole moment of KRb $d = 0.566 \, ea_0$~\cite{ni2008high}. For weak field, the eigenstates of this Hamiltonian are the rotational eigenstates $\ket{\mathcal N, \mathcal N_Z}$ with $\hat N^2 \ket{\mathcal N, \mathcal N_Z} = \mathcal N \qty(\mathcal N+1)\ket{\mathcal N, \mathcal N_Z}$, where the electric field lifts the degeneracy between different $\mathcal N_Z$.

For $M=0$ states, the dipoles are oriented parallel to the electric field, and the angular dependence of the dipole-dipole interactions is
\begin{align}
    V_\mathrm{dd}(\vec r) = \frac{1 - 3\cos^2(\theta)}{r^3}\, ,
\end{align}
where $\theta$ is the angle of molecule separation relative to the electric field axis and $r = \abs{\vec r}$. In second quantization, the interaction Hamiltonian reads
\begin{align}
    \hat H_\mathrm{dd} = \frac{1}{2} \int d^3r \int d^3r' \, V_\mathrm{dd}(\vec r - \vec r')
    \qty{\qty[\sum_{mm'} \mu_m \mu_{m'} \hat \psi_m^\dagger(\vec r) \hat \psi_{m'}^\dagger(\vec r') \hat \psi_{m'}(\vec r') \hat \psi_m(\vec r)] +
    \mu_{\uparrow\downarrow}^2 \qty[\hat \psi_\uparrow^\dagger(\vec r) \hat \psi_\downarrow^\dagger(\vec r') \hat \psi_\uparrow(\vec r') \hat \psi_\downarrow(\vec r) + h.c.]} \, , \label{eq:sup_hdd}
\end{align}
with $(m, m' = \uparrow, \downarrow)$, field operators $\hat \psi$, the transition dipole moment $\mu_{\uparrow\downarrow} = \bra{\uparrow} \hat d_0 \ket{\downarrow}$, and permanent dipole moments $\mu_m$ of state $m$. At zero field, $\mu_{\uparrow\downarrow} = d/\sqrt{3}$ and $\mu_m = 0$.

When expanding Eq.~\eqref{eq:sup_hdd} into Wannier orbitals that are strongly localized, only those terms with $\hat \psi^\dagger_\sigma(\vec r)$ and $\hat \psi_{\sigma'}(\vec r)$ on the same lattice site contribute. The dipole contribution to the on-site interactions is then given by the on-site integral
\begin{align}
    U_\mathrm{dd} = \frac{1}{2} \int d^3 r d^3 r' \, \mu_{\uparrow\downarrow}^2 V_\mathrm{dd}(\vec r - \vec r') \qty[\psi_\uparrow^*(\vec r) \psi_\downarrow(\vec r) \psi_\downarrow^*(\vec r') \psi_\uparrow(\vec r') + c.c.] \, .
\end{align}
Numerical values for $U_\mathrm{dd}$ are given in Tab.~\ref{tab:lattice_params}.

In order to estimate the nearest neighbor dipole interactions, we again assume strongly localized orbitals such that we can replace $V_\mathrm{dd}(\vec r - \vec r') \approx V_\mathrm{dd}(\vec a)$. We then find
\begin{align}
    \hat H_\mathrm{dip} = \sum_{i,j} \frac{1}{2} V_\mathrm{dd}(\vec a) \mu_{\uparrow\downarrow}^2 \qty(\hat s^+_i \hat s^-_j + \hat s^-_i \hat s^+_j) \iint d^3r d^3r' \, \psi_\uparrow^*(\vec r) \psi_\downarrow(\vec r) \psi_\downarrow^*(\vec r') \psi_\uparrow(\vec r') \, ,
\end{align}
where the integral becomes 1 for $\psi_\uparrow(\vec r) = \psi_\downarrow(\vec r)$, and is less than 1 otherwise. When the extent of the Wannier orbitals approaches the lattice spacing further corrections need to be taken into account~\cite{wall2017microscopic}. Using $0 \leq \cos^2(\theta) \leq 1$ and $a = 532$nm, we find $-220\text{Hz} \leq V_\perp \leq 110$Hz.

\subsubsection*{Decay rate}
In order to derive an expression for the decay rate, we adapt the derivation of Ref.~\cite{garcia2009dissipation} to fermions. The idea is to first derive an effective Lindbladian description of the free space scenario and then integrate over a lattice site, analogous to elastic $s$-wave scattering/contact interactions. In a 3D gas with inelastic scattering, the evolution of the number of particles in state $\ket \uparrow$ is determined by
\begin{align}
    \partial_t n_\uparrow &= - n_\uparrow \sigma v \, , \\
    v &= \frac{\hbar k}{m}\, .
\end{align}
Here, $v$ is the velocity, $m$ is the mass, $\hbar k$ is the momentum, and $\sigma$ is the inelastic scattering cross section with particles in $\ket\downarrow$. It depends on the complex scattering length $a_\mathrm{sc}$ as~\cite{hutson2007feshbach}
\begin{align}
    \sigma &= n_\downarrow \times \frac{-4\pi \Im (a_\mathrm{sc})}{k[1 + k^2 \abs{a_\mathrm{sc}}^2 + 2k\Im(a_\mathrm{sc})]} \, .
\end{align}
Thus, in the limit where particles have small momentum compared to the scattering length $k\abs{a_\mathrm{sc}} \ll 1$
\begin{align}
    \partial_t n_\uparrow = \frac{4 \pi \hbar \Im(a_\mathrm{sc})}{m} \times n_\uparrow n_\downarrow \, . \label{eq:sup_dnuprate}
\end{align}
For KRb, in the paper we use the universal $s$-wave scattering parameters~\cite{idziaszek2010universal}, which have been experimentally confirmed by measurements of loss rates for molecules in two different hyperfine states~\cite{miranda2011controlling}. For rotational states, measurements in a 3D lattice observe a factor of 5 larger loss rate~\cite{yan2013observation}, which require the inclusion of higher bands when computing the dynamics~\cite{zhu2014suppressing}. Nevertheless, the renormalized loss rate parameters were similar to the ones obtained for nuclear spin states. Therefore, we use the latter in this paper.

\bigskip

We match this evolution to a master equation with Lindblad operator
\begin{align}
    \hat L_{\vec r} = \sqrt{\frac{\Gamma_\mathrm{3D}}{2}} \hat \psi_\uparrow(\vec r) \hat \psi_\downarrow(\vec r) \, ,
\end{align}
where $\psi_\sigma(\vec r)$ is the fermionic annihilation operator for a molecule in state $\sigma$ at position $\vec r$, and $\Gamma_\mathrm{3D}$ is the free space decay rate which we want to determine.

We can also calculate the time evolution of the population of $\ket\uparrow$ from the master equation as
\begin{align}
    \partial_t \langle n_\uparrow \rangle &=
    \frac{\Gamma_\mathrm{3D}}{2} \left\langle \int d^3r 
    \qty[2 \hat \psi_\downarrow^\dagger(\vec r) \hat \psi_\uparrow^\dagger(\vec r) \hat n_\uparrow \hat \psi_\uparrow(\vec r) \hat \psi_\downarrow(\vec r) 
    - \hat n_\uparrow \hat \psi_\downarrow^\dagger(\vec r) \hat \psi_\uparrow^\dagger(\vec r) \hat \psi_\uparrow(\vec r) \hat \psi_\downarrow(\vec r)
    - \hat \psi_\downarrow^\dagger(\vec r) \hat \psi_\uparrow^\dagger(\vec r) \hat \psi_\uparrow(\vec r) \hat \psi_\downarrow(\vec r) \hat n_\uparrow] \right\rangle \, .
\end{align}
To evaluate this integral, we use $\hat n_\uparrow = \int d^3r' \hat \psi_\uparrow^\dagger(\vec r') \hat \psi_\uparrow(\vec r')$. For $\vec r \neq \vec r'$, the terms in the integral cancel. For $\vec r = \vec r'$, the first term vanishes due to $\hat \psi_\uparrow^\dagger(\vec r)^{2} = 0$. For the other terms, we use the fermionic identity $\hat \psi^\dagger(\vec r) \hat \psi(\vec r) \hat \psi^\dagger(\vec r) \hat \psi (\vec r) = \hat \psi^\dagger(\vec r) \hat \psi (\vec r)$. We are left with
\begin{align}
    \partial_t \langle n_\uparrow \rangle &=
    -\Gamma_\mathrm{3D} \left\langle \int d^3 r \,
    \hat\psi_\downarrow^\dagger(\vec r) \hat\psi_\downarrow(\vec r) \hat\psi_\uparrow^\dagger(\vec r) \hat\psi_\uparrow(\vec r)
    \right\rangle\, . \label{eq:sup_dnupmaster}
\end{align}

From matching Eqs.~\eqref{eq:sup_dnuprate} and \eqref{eq:sup_dnupmaster}, we can easily extract
\begin{align}
    \Gamma_\mathrm{3D} = \frac{-4\pi \hbar \Im(a_\mathrm{sc})}{m} \, .
\end{align}
By integrating Eq.~\eqref{eq:sup_dnupmaster} over the Wannier functions, we can derive the decay rate
\begin{align}
    \Gamma = \frac{-4\pi \hbar \Im(a_\mathrm{sc})}{m} \int  d^3r \, \abs{\psi(\vec r)}^4\, .
\end{align}
Finally, for strong decay rates, molecules obey the universal condition $-\Im(a_\mathrm{sc}) = \Re(a_\mathrm{sc})$, leading to $\Gamma = U_\mathrm{contact}/\hbar$, which is given in Tab.~\ref{tab:lattice_params}.

\section{Interaction Range}
\begin{figure}
    \centering
    \includegraphics[width=0.5\textwidth]{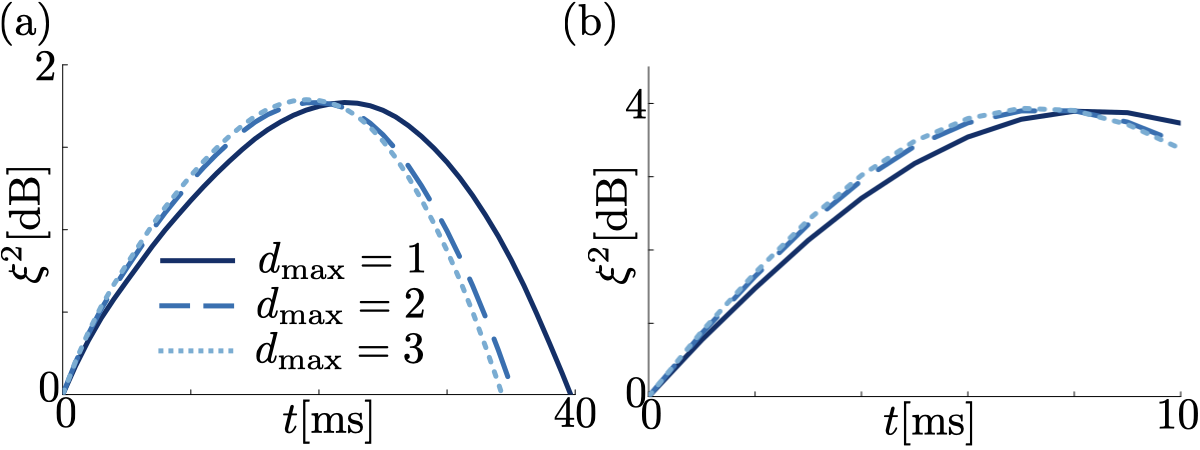}
    \caption{Truncation of interaction radius. Squeezing evolution for different interaction truncation radii $d_\mathrm{max}$, i.e.~dipolar interactions between sites $i$ and $j$ are set to zero for $\abs{i-j} > d_\mathrm{max}$. We simulate coherent dynamics on $L=20$ lattice sites with (a) $N=4$ and (b) $N=16$ starting from random 32 and 8 different initial states, respectively (identical initial states for different $d_\mathrm{max}$). Changes in peak height are smaller than statistical or MPS truncation errors (not shown). Other parameters: $J/h = 153$Hz, $U/h = 434$Hz, $\Gamma = 0$, $V_\perp/h = 40$Hz. $\chi = 256$ for $N=4$ and $\chi = 1024$ for $N=16$.}
    \label{fig:intrange}
\end{figure}

In Fig.~\ref{fig:intrange}, we show the effect of including beyond nearest neighbor terms in a system without losses averaged over different initial states. We find that the main contribution is a re-scaling of time, whereas changes to the maximal squeezing are smaller than simulation errors. We attribute this to two competing effects of the long-range tails on squeezing. On the one hand, the connectivity is increased, which is generally expected to increase squeezing. On the other hand, there is an effective increase in interaction strength by a factor $\sum_j 1/j^3 \approx 1.2$. As discussed in the main text, increasing the interaction strength speeds up the dynamics, but reduces squeezing. The resulting effects on squeezing approximately cancel each other, such that only the speed-up remains.

\section{MPS Simulation}

\subsection{Infinite Time Evolving Block Decimation}
In order to represent the full density matrix as an MPS directly in the thermodynamic limit, we use the translation invariance of the state~\cite{orus2008infinite}. In particular, we write the density matrix as
\begin{align}
    \hat \rho = \sum_{\{i_n\}} \sum_{\{\alpha_n\}} \prod_n \Gamma^{[n]}_{i_n,\alpha_n,\alpha_{n+1}} \lambda^{[n]}_{\alpha_n} \bigotimes_n \hat e_{i_n} \, , \label{eq:sup_imps}
\end{align}
where the $i_n$ run over the physical dimension, and the $\alpha_n$ run over the virtual bond dimension and are truncated such that $1 \leq \alpha_n \leq \chi$. The $\hat e_{i_n}$ form a basis set of the local density matrices (i.e.~$1\leq i_n \leq 16$ for four possible states on each site), for which we here choose the generalized Gell-Mann matrices~\cite{schachenmayer_privcomm}. For a translation invariant state, also the $\Gamma$ and $\lambda$ tensors are translation invariant, i.e.~$\Gamma^{[n]} = \Gamma^{[n+2]}$ and $\lambda^{[n]} = \lambda^{[n+2]}$, and we only keep two $\Gamma$ and $\lambda$ tensors. In principle, $\Gamma$ and $\lambda$ are translation invariant by one site, but during the time evolution this translation invariance is intermittently broken and then restored, making all four tensors necessary.

To time evolve the state, we use an infinite time-evolving block decimation algorithm, i.e.~a Trotter decomposition of the Hamiltonian into two site gates. However, a naive implementation of this algorithm for density matrices would destroy the orthogonality of the MPS. This is fixed by re-orthogonalizing the MPS after every gate application~\cite{orus2008infinite}. For all simulations except the spin model simulation, we use a fourth order Trotter decomposition~\cite{sornborger1999higher}
\begin{align}
    (1)(1)^T(1)(-2)^T(1)(1)(1)(1)(1)^T(1)(1)^T(1)^T(1)^T(1)^T(-2)(1)^T(1)(1)^T \, . \label{eq:sup_trotter4}
\end{align}
Here, the number in brackets indicates the length of the timestep, negative numbers for evolution backwards, and the superscript $T$ indicates that we apply the gates in transposed order.
For the spin model simulation, we use a smaller time step for which the second order trotterization $(1)(1)^T$ suffices.

\begin{figure}
    \centering
    \includegraphics[width=0.8\textwidth]{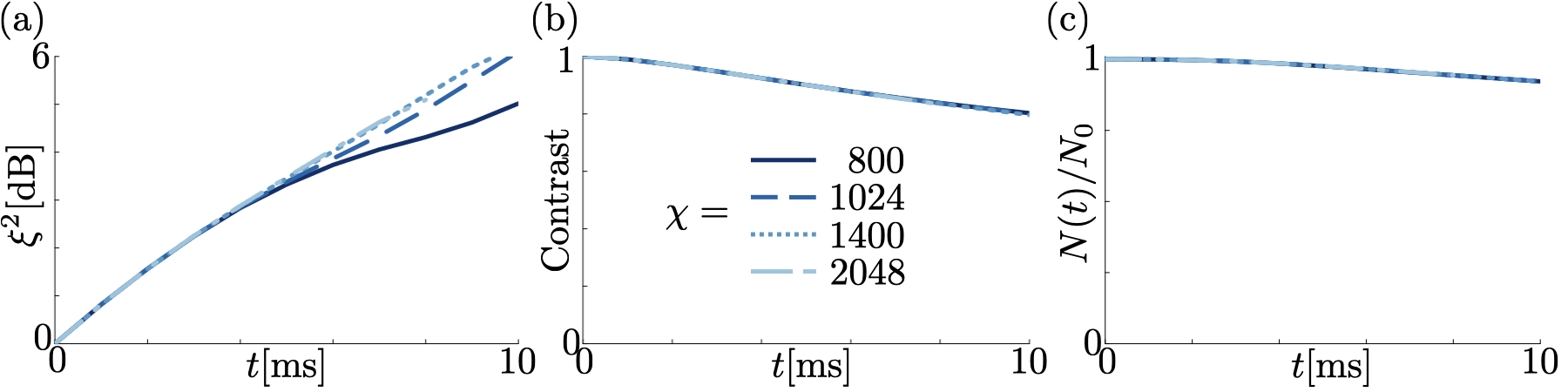}
    \caption{Convergence with bond dimension $\chi$ for the shallowest lattice with $V_X=V_Y=V_Z = 3E_R $, for a filling fraction $f=0.8$. Only ``worse'' case is $f=1$ for the same lattice depth. For $\chi = 2048$, the squeezing is converged until $t=8$ms. Contrast and molecule number are converged for all cases shown.}
    \label{fig:chi-conv}
\end{figure}

\begin{table}
    \centering
    \begin{tabular}{c|cccccc}
        \backslashbox{$(J/U)$}{$f$} & 0.1 & 0.2 & 0.4 & 0.6 & 0.8 & 1 \\
        \hline
        1.5      & 1024 & 1024 & 1400 & 2048 & 2048 (X) & 2048 (X) \\
        0.3      & 1024 & 1024 & 1400 & 1400 & 1400 & 2048 (X) \\
        0.15     & 1024 & 1024 & 1024 & 1400 & 1400 & 1400 \\
        $<10^{-3}$ & 1024 & 1024 & 1024 & 1024 & 1024 & 1024 \\
    \end{tabular}
    \caption{Bond dimension $\chi$ used for different filling fractions $f$ and lattice depths indicated by $J/U$ (see Tab.~\ref{tab:lattice_params}). Entries marked by (X) are not conclusively converged until $t=10$ms. This does not affect any results shown in the paper.}
    \label{tab:bond_dim}
\end{table}

\subsection{Convergence}
In order to ensure convergence of our simulation, we successively decrease the timestep $\Delta t$ and increase the bond dimension until convergence is reached. In practice, we halve the timestep and increase the bond dimension in steps of $\chi = 512,750/800,1024,1400,2048$, until the curves overlap. For the time step, we have confirmed for all results with $f=0.1$ and for the shallowest lattice with $f=0.8$ that a timestep $\Delta t = 1$ms [time for a full Trotter sequence Eq.~\eqref{eq:sup_trotter4}] is sufficient, independent of the system parameters. For the second order decomposition $\Delta t = 25\mu$s is needed. The procedure to determine the bond dimension is illustrated in Fig.~\ref{fig:chi-conv} at fixed $\Delta t = 1$ms in one of the worst case scenarios. The bond dimension strongly depends on the system parameters. For Figs.~1, 2, and 4, the bond dimensions used are shown in Tab.~\ref{tab:bond_dim}. In Fig.~3, we find that $\chi = 4096$ for the full Fermi-Hubbard simulation and $\chi = 512$ for the spin model are sufficient.

\subsection{Computing Correlations}
For fully coherent dynamics (results in Fig.~3), we compute squeezing by computing all correlations including up to 80 neighbors in each direction and compute the squeezing from there.
For the incoherent dynamics, we directly compute the infinite correlations $\sum_{n > 0} \langle \hat O^{(1)}_0 \hat O^{(2)}_n \rangle $, $\hat O^{(i)} \in \{ \hat s^x ,\hat s^y, \hat s^z\}$ with an approach closely related to transfer matrices. We can write $\sum_{n > 0}^L \hat O^{(2)}_n$ as an $L$-site matrix product operator (MPO), that is in the form of Eq.~\eqref{eq:sup_imps} with $\lambda^{[n]} = \mathbb 1$~\cite{schollwock2011density}.
The $\Gamma$ tensors are
\begin{align*}
    \Gamma^{[1]} = 
    \begin{pmatrix}
    \hat O^{(2)}_1 & \mathbb 1
    \end{pmatrix}
    ,\quad
    \Gamma^{[L]} = 
    \begin{pmatrix}
    \mathbb 1 \\
    \hat O^{(2)}_N
    \end{pmatrix}
    ,\quad
    \Gamma^{[n]} = 
    \begin{pmatrix}
    \mathbb 1 & 0 \\
    \hat O^{(2)}_n & \mathbb 1
    \end{pmatrix}
    , 1 < n < L,
\end{align*}
where the matrix dimensions indicate the bond dimension, and the operators in the matrix elements are vectorized to give the physical dimension.

In order to compute $\Tr[\hat \rho \hat O^{(1)}_0 \sum_{n > 0}^L \hat O^{(2)}_n]$, we contract their MPS/MPO representations along the physical dimension. The lattice sites with no operator acting on them are contracted with the identity. The contraction over the indices $1 < n < L-1$ can be written as the $[(L-2)/2]$th power of the matrix
\begin{align}
    M^{\alpha_n\beta_n}_{\alpha_{n+2}\beta_{n+2}} = \sum_{i_n, i_{n+1}} \sum_{\alpha_{n+1}} \sum_{\beta_{n+1}}
    \Gamma^{[n](\rho)}_{i_n,\alpha_n,\alpha_{n+1}} \lambda^{[n](\rho)}_{\alpha_n}
    \Gamma^{[n+1](\rho)}_{i_{n+1},\alpha_{n+1},\alpha_{n+2}} \lambda^{[n+1](\rho)}_{\alpha_{n+1}}
    \Gamma^{[n](O)}_{i_n,\beta_n,\beta_{n+1}} \lambda^{[n](O)}_{\beta_n}
    \Gamma^{[n+1](O)}_{i_{n+1},\beta_{n+1},\beta_{n+2}} \lambda^{[n+1](O)}_{\beta_{n+1}} \label{eq:sup_mpocontract}
\end{align}
where the superscript $\rho$ or $O$ indicate from which representation the operators were taken, and $\{\alpha_n,\beta_n\}$ is interpreted as a combined left index, while $\{\alpha_{n+2},\beta_{n+2}\}$ is interpreted as a combined right index. This power can be easily computed by diagonalizing the matrix $M$. Using the diagonalized form, it is straightforward to take the limit $L \rightarrow \infty$.

The sites $n<0$ and $n>L$ are contracted with the vectorized identity matrix. This contraction can also be written in the form of Eq.~\eqref{eq:sup_mpocontract}, where the dimension of the $\beta_n$ indices is 1. Again, the infinite power of this matrix is easily computed by diagonalizing it. We note that we only find eigenvalues $\leq 1$ for the matrices, ensuring that the limit $L\rightarrow \infty$ is well defined.

\section{Spin Model Derivation}
For unit filling $f=1$ and no losses $\Gamma = 0$, the spin model can be derived by adiabatically eliminating all states with one or more doubly occupied sites. Using the formalism of Ref.~\cite{reiter2012effective} with $\hat V = -\sum_{j,\sigma} J (\hat b_{j,\sigma}^\dagger \hat b_{j+1,\sigma} + h.c.)$, $\hat H_e = U \sum_j \hat n_{j\uparrow} \hat n_{j\downarrow }$ and $\hat H_g = \hat H_\mathrm{dip}$, it is straightforward to derive the given terms. Note that the $\vec s_j \vec s_{j+1}$ term arises due to fermionic statistics, which only allows molecules with opposite spin states to tunnel to the same site. A similar term would also arise for bosons, however with opposite sign. In addition, a constant energy shift was dropped.

\section{Molecule Number Decay}

\begin{figure}
    \centering
    \includegraphics[width=\textwidth]{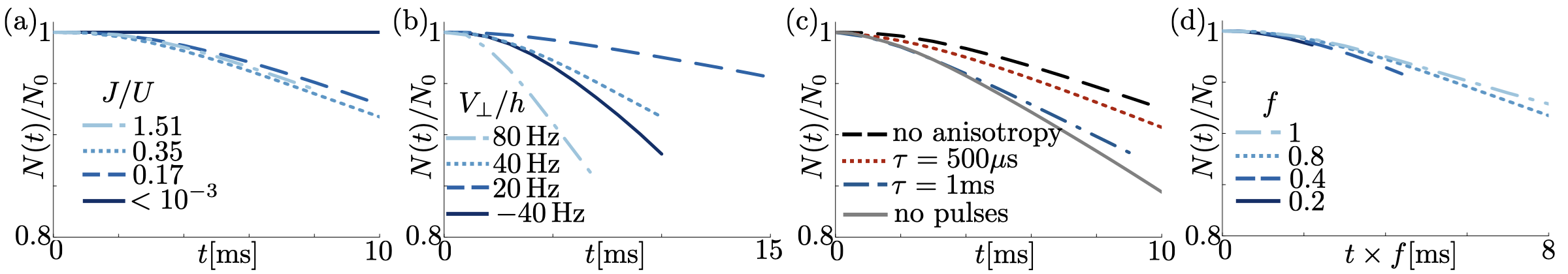}
    \caption{Molecule loss for the different cases discussed in the paper. (a) corresponds to Fig.~2(a/b); (b) corresponds to Fig.~2(c/d); (c) corresponds to Fig.~4(a); (d) corresponds to Fig.~4(b).}
    \label{fig:num_dec}
\end{figure}

The molecule loss for the time traces analyzed in the main paper is shown in Fig.~\ref{fig:num_dec}. In all cases, the molecule number at peak squeezing is $\gtrsim 90\%$ of the initially present molecules. In panel (a), for the deepest lattice with $J/U < 10^{-3}$ the molecules do not decay at all. For the other lattice depths, we find some losses, with the loss rate being approximately independent of the lattice depth. In the other panels losses are slowest when contrast and squeezing are largest (compare corresponding cases in the main paper). This demonstrates a combination of two effects suppressing molecule decay. Especially for deep lattices, creating doubly occupied sites is energetically forbidden and Zeno-suppressed. In addition, when the contrast is large, doubly occupied sites cannot be formed due to the Pauli exclusion principle, thus preventing losses.

\section{Dephasing}

\begin{figure}
    \centering
    \includegraphics[width=0.8\textwidth]{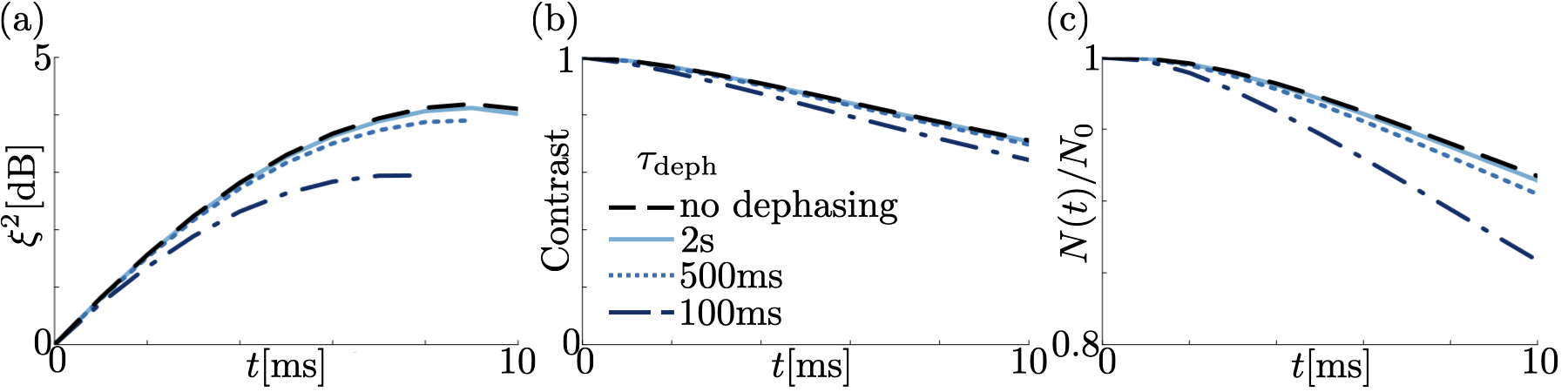}
    \caption{Evolution under dephasing of (a) squeezing, (b) contrast, and (c) molecule number for different spin coherence times $\tau_\mathrm{deph}$. Parameters: $J/h = 153$Hz, $U/h = 434$Hz, $\Gamma = 2\pi \times 512$s$^{-1}$, $V_\perp/h = 40$Hz.}
    \label{fig:dephasing}
\end{figure}

In Fig.~\ref{fig:dephasing}, we consider the effect of dephasing on the system dynamics. Here, we model dephasing as white noise on the two different spin states, described by Lindblad operators $\hat L_j = \sqrt{\Gamma_\mathrm{deph}} \hat s^z_j$. For spin coherence times of $\tau_\mathrm{deph} = 1/\Gamma_\mathrm{deph} = 100$ms, we find a significant reduction in squeezing. In contrast, for spin coherence times of $500$ms, squeezing is only slightly reduced, and for $\tau_\mathrm{deph} = 2$s, changes induced by dephasing are barely visible. Since recent experiments have reported interaction limited coherence times $> 400$ms, we can assume that the real coherence times are even longer, such that additional dephasing does not need to affect the spin squeezing.

\end{widetext}

\end{document}